\documentclass[runningheads]{llncs}

\usepackage[T1]{fontenc} % Ensure correct character encoding
\usepackage{graphicx} % For including figures
\usepackage{hyperref} % For clickable links
\usepackage{biblatex} % For bibliography management
\usepackage{booktabs}
\begin{document}
\title{Enhancing Large Language Models with Faster Code Preprocessing for Vulnerability Detection}
\author{José Gonçalves\orcidID{0009-0004-1038-8384} \and
Miguel Silva\orcidID{0009-0008-6630-9939} \and
Eva Maia\orcidID{0000-0002-8075-531X}\and
Isabel Praça\orcidID{0000-0002-2519-9859}}

\authorrunning{J. Gonçalves et al.}% First names are abbreviated in the running head.
\titlerunning{Enhancing Large Language Models}
\institute{GECAD, ISEP, Polytechnic of Porto, rua Dr. António Bernardino de Almeida, 4249-015 Porto, Portugal \email{\{jpsgs,mdgsa,egm,icp\}@isep.ipp.pt}}

\maketitle              % typeset the header of the contribution
\begin{abstract}
The application of Artificial Intelligence has become a powerful approach to detecting software vulnerabilities. However, effective vulnerability detection relies on accurately capturing the semantic structure of code and its contextual relationships. Given that the same functionality can be implemented in various forms, a preprocessing tool that standardizes code representation is important. This tool must be efficient, adaptable across programming languages, and capable of supporting new transformations. To address this challenge, we build on the existing SCoPE framework and introduce SCoPE2, an enhanced version with improved performance. We compare both versions in terms of processing time and memory usage and evaluate their impact on a Large Language Model (LLM) for vulnerability detection. Our results show a 97.3\% reduction in processing time with SCoPE2, along with an improved F1-score for the LLM, solely due to the refined preprocessing approach. 

\keywords{Artificial Intelligence \and Code Processing \and Efficiency  \and Large Language Model \and Vulnerability Detection.}
\end{abstract}

\section{Introduction}
%\subsection{A Subsection Sample}

Software is integral to modern life, powering everything from household appliances to critical infrastructures.  This ubiquity presents significant challenges, because even minor bugs or vulnerabilities can lead to severe disruptions, especially when exploited by malicious actors. In this context, Artificial Intelligence (AI) has emerged as a critical tool, supporting tasks ranging from code analysis to vulnerability detection and automated code generation. However, code generation and analysis are inherently complex tasks because they involve processing sequences of tokens, each of which carries a unique semantic value \cite{10531717}. Unlike natural language, where some variability in word order or sentence structure is acceptable, source code adheres to strict syntactic rules, and even small structural changes can significantly affect its functional behavior \cite{10.1145/3428230}. Despite these rigid syntactic constraints, the semantics of code allow for remarkable flexibility, since the same functionality can often be implemented in multiple ways, expanding the range of possible implementations and, consequently, the challenges of analysis \cite{Yu2023}.

For an AI model to accurately analyze code, it must deeply understand the semantic role of each token and how these tokens interact within the broader program context. However, even state-of-the-art AI models often struggle with this level of comprehension \cite{10.1145/3630010}. Research has shown that AI models for Software Vulnerability Detection (SVD) frequently rely on superficial elements, such as programmer-defined variable names, rather than deeper structural patterns. To further investigate this AI-model understanding limitation, some studies \cite{10.1145/3611643.3616356} apply transformations to source code to generate new samples that can be used with multiple purposes, namely to improve the generalization of the model. However, no standardized tool exists for this purpose, as current implementations are often constrained by the target programming language, software extensibility, or the original objectives of their developers. As a result, multiple researchers create similar transformation tools, leading to fragmentation and limiting comparability across studies. A significant step toward a universal solution for code transformation was introduced in a previous work with Source Code Processing Engine (SCoPE) \cite{gonçalves2024scopeevaluatingllmssoftware}, a tool that enables transformations, such as variable renaming, to be applied to source code. While promising, SCoPE has notable limitations that pose challenges to its broader adoption.

This work addresses the lack of a standardized tool for applying transformations to source code by improving and extending the existing SCoPE tool. The new version overcomes the main limitations of its predecessor while introducing a highly flexible framework that simplifies the application of source code transformations. In this study, the tool is used to preprocess data for fine-tuning a large language model (LLM) for SVD, facilitating source code preparation for model training or inference. The objective is to assess whether the updated version of SCoPE, which supports a wider range of code samples, including those containing macros, improves model performance compared to its previous version. In summary, the main contributions of this paper are as follows:

\begin{itemize}
    \item Update of the SCoPE framework. The new version is intended to lay the foundations for the adoption of a common platform for the application of transformations to source code. The code will be available on Github~\footnote{https://github.com/jp2425/SCoPE2}.
    \item Fine-tuning of Large Language Model Meta AI (LLaMA) 3.2, a compact yet powerful model released by Meta AI, for vulnerability detection. The training data was processed using the latest version of SCoPE, with results compared against those obtained from the original version.
\end{itemize}

The structure of this study is as follows: Section \ref{sec:sota} presents a literature review on the applications of LLMs for SVD, challenges in this field, the role of code transformations, and the need for further research. Section \ref{sec:scope} introduces SCoPE2 and compares its performance to the previous version. Section \ref{sec:exp_setup} details the experimental setup, including the dataset, data preprocessing, and the LLaMA 3.2 model. The results are presented and discussed in Section \ref{sec:results}. Finally, Section \ref{sec:conclusion} summarizes the main findings and outlines potential directions for future research.

\section{Related Work}
\label{sec:sota}
As software is a significant component of most devices, vulnerabilities pose a significant security threat, potentially exposing systems to exploitation by malicious actors. A vulnerability can compromise an entire system, leading to data breaches and security incidents, making SVD a core practice for maintaining high security standards \cite{KALOUPTSOGLOU2023107303}.

Traditional SVD methods rely heavily on manual inspection to verify the presence of vulnerabilities. This process is usually time-consuming and labor-intensive, requiring skilled developers to meticulously analyze code and identify potential vulnerabilities, while also being able to understand the rationale behind recent code changes \cite{LOMIO2022111283}. Despite their effectiveness in some instances, traditional approaches often struggle with scalability and have a high false positive rate \cite{10172650}. As software projects expand in size and complexity, these limitations become increasingly apparent, making it challenging to keep up with evolving security threats. Consequently, extensive research has focused on AI-driven techniques to enhance and automate vulnerability detection, addressing the shortcomings of traditional methods. \cite{10172650}.

The recent widespread adoption of LLMs has further demonstrated their potential as highly effective tools for SVD \cite{LU2024112031,10908394}. These models have shown strong performance in language-related tasks \cite{10301302}, particularly in identifying relationships within textual sequences \cite{10774025}. With their ability to understand code \cite{10879492}, LLMs can uncover coding patterns and distinguish between secure and vulnerable code \cite{10774025}, making them a promising approach for automated vulnerability detection \cite{10114536394763639762}.

The limited token processing capacity of LLMs is an important factor in their application. To address this issue, source code can be modified to reduce its size. However, achieving an optimal representation is challenging, especially in SVD, where multiple transformations can produce functionally equivalent but structurally different versions of the same code. To address this, the SCoPE framework \cite{gonçalves2024scopeevaluatingllmssoftware} has been introduced, offering various preprocessing techniques specifically for C/C++ code. This framework can apply function-level transformations to ensure standardized code representations. These transformations include shortening variable and function names, removing comments and error handling strings, as well as normalizing whitespaces. By generalizing programmer-defined elements, such as variable names, SCoPE can be also used to verify the existence of duplicate entries, which is particularly relevant when datasets are automatically generated by analyzing code differences.
When a function undergoes minor changes, such as a different function name, it may appear multiple times in the dataset while retaining the same functionality. However, its labeling is not always consistent, as shown in a previous work \cite{gonçalves2025evaluatingllama32software}. 

Although SCoPE is an important tool for standardizing source code for AI-based approaches, it has some limitations. Since it relies on \texttt{Antlr4} \footnote{https://github.com/antlr/antlr4} %\cite{parr2013antlr} 
parser, it lacks flexibility, which limits its adaptability to new functionalities, transformations, and programming languages. In addition, its inability to handle code containing macros and its inefficiency when applied to process large datasets, make it impractical for broader applications. Processing large datasets with \texttt{Antlr4} is time-consuming, further highlighting the need for an optimized solution.

Overall, while SCoPE has demonstrated its effectiveness in producing standardized code representations, its constraints, particularly in terms of flexibility, scalability, and processing speed, underline the need for a more advanced and optimized version. As research in this area continues to evolve, the development of an improved framework capable of overcoming these limitations is imperative.

\section{SCoPE2}
\label{sec:scope}

The SCoPE2 framework is an improved version of the original SCoPE framework, addressing scalability challenges posed by \texttt{Antlr4}, where different grammars required entirely distinct code visitors to traverse their respective parse trees. As a result, extending support for languages beyond C/C++ in the original SCoPE framework incurred a significant cost.

Therefore, to address the limitations of the original implementation, SCoPE2 was created using a new parser. \texttt{TreeSitter} \footnote{https://github.com/tree-sitter/tree-sitter} was chosen as an alternative to \texttt{Antlr4}, as it is more efficient, scalable, and does not have the constraints associated with handling macros in code. The implementation was guided by two primary principles: 
\begin{itemize} 
    \item \textbf{Efficiency}: The new implementation was designed to be as efficient as possible. This principle influenced several design decisions, such as the consolidation of all \texttt{TreeSitter} queries into a single transformation query to avoid redundant parsing rounds, one for each transformation that needed to extract information from the parser. 
    \item \textbf{Flexibility}: The framework is intended to support the addition of new transformations by users. It was designed to be extensible and not constrained by the original framework's scope or any specific programming language. 
\end{itemize}

In an initial iteration, all transformations supported by the original SCoPE framework were ported to SCoPE2. These transformations include string generalization, variable name generalization, function name generalization, comment removal, and code tokenization. These transformations are generic in nature, and to apply them to other languages, the only requirement is to provide the \texttt{TreeSitter} queries for the relevant language elements used in the transformations. 

The usage of SCoPE2 is conceptually straightforward: the user provides a configuration for the transformations (or uses the default configuration if none is provided), the code to be processed, a list of selected transformations, and a class indicating the desired final representation of the code. Currently, SCoPE2 supports two code representations as the framework's output: text and token array, two representations common for AI training. Further details on this process are provided on the project’s GitHub repository.

\section{Experimental Setup}
\label{sec:exp_setup}
The study conducted in this work explores the usage of SCoPE2 to create a more generic representation of code. The experiments were conducted on Kaggle Notebooks, with a Nvidia P100 GPU, with 16GB of VRAM.

\subsection{Dataset and Data Pre-Processing}

DiverseVul \cite{Chen2023} is a comprehensive vulnerability dataset containing 18,945 vulnerable C/C++ functions spanning 150 Common Weakness Enumerations. It is built from real-world data extracted from open-source repositories, compiling a large collection of both vulnerable and non-vulnerable C/C++ functions sourced from 797 projects. RDiverseVul \cite{goncalves_2025_15051277} is an improved version of DiverseVul, with erroneous data identified using SCoPE and subsequently removed. Specifically, code snippets that were structurally similar but labeled inconsistently due to differences only in function names were eliminated. A summary of the dataset's key features is provided in Table \ref{tab:dataset_features}.

\begin{table}[ht]
\caption{RDiverseVul Dataset Features.}\label{tab:dataset_features}%
\centering
\begin{tabular}{@{}ll@{}}
\toprule
\textbf{Feature} & \textbf{Description} \\
\midrule
func & Contains the C/C++ function code. \\
target & Is the function vulnerable or not.  \\
cwe & List of respective CWEs present in the function. \\
project & Project from where the function was extracted. \\
commit\_id & Identifier of the commit. \\
size & Size of the function. \\
message & Commit message. \\
\bottomrule
\end{tabular}
\end{table}

LLaMA 3.2, a model with 1.23 billion parameters, was adapted for SVD. Originally developed for sequence-to-sequence generation and pre-trained on a large multilingual dataset for general language tasks, it was fine-tuned for binary classification of C/C++ code samples, identifying them as either vulnerable or not. To address computational constraints, low-rank adaptation (LoRA) was employed during fine-tuning. As a Parameter-Efficient Fine-Tuning (PEFT) technique, LoRA significantly reduces the number of trainable parameters, cutting training time while maintaining model performance \cite{Razuvayevskaya2023Comparison}.

The hyperparameters used in this study were based on the best hyperparameters found in a previous study \cite{gonçalves2025evaluatingllama32software}. To ensure a fair comparison, this work also uses a 10000 subset of the RDiverseVul dataset and uses the same proportions for the split into training, evaluation and test subsets.

\subsection{Performance Analysis}

Following the development of SCoPE2, a comparative analysis was conducted, focusing on two key performance metrics: processing time and memory usage during execution. These metrics allow to evaluate the scalability and efficiency of the framework, especially when dealing with large codebases and resource-constrained environments.

Both SCoPE versions were tested using a simple script that leveraged the \texttt{pandarallel} library \footnote{https://github.com/nalepae/pandarallel} %\cite{nalepa_pandarallel} 
to process 1000 entries from RDiverseVul \cite{gonçalves2025evaluatingllama32software}. This library allows for parallel operations on pandas dataframes, using multiple CPU cores on the same operation. Four threads were used for processing. The same subset of the RDiverseVul dataset was used for the entire experiment. The selected transformations for both versions were function name and variable name generalization, as well as comment removal.

Memory usage during execution was monitored using the \texttt{memory\_profiler} tool ~\footnote{\url{https://github.com/pythonprofilers/memory_profiler}}, while execution time was measured using the Linux \texttt{time} utility. To ensure a fair comparison, the original SCoPE version was configured to utilize \texttt{Antlr4’s} ability to recover from code errors, a feature that is handled automatically by \texttt{TreeSitter}. SCoPE2 and SCoPE were configured to return the modified code in plain text.

The execution time shows a significant difference between the two implementations. The original SCoPE framework required 3 minutes and 42 seconds to process 1000 entries, whereas SCoPE2 completed the same task in just 6 seconds. This means SCoPE2's execution time is only 2.7\% of the original implementation, highlighting a substantial performance improvement. Table \ref{tab:scope_comparison} presents the results of the tests.

\begin{table}[ht]
    \centering
    \caption{Performance Comparison of SCoPE Versions.}
    \label{tab:scope_comparison}
    \begin{tabular}{l@{\hskip 0.5cm}l@{\hskip 0.5cm}l}
    \hline
    \textbf{SCoPE Version} &  \textbf{Execution Time} & \textbf{Peak Memory Usage} \\ \hline
    SCoPE & 3m 42s & 84.9 MB \\
    SCoPE2 &  6s & 79.2 MB \\ \hline
    \end{tabular}
\end{table}

In terms of memory usage, the difference between the two versions was less pronounced. The maximum memory usage for the script using SCoPE2 was 79.2 MB, with most of the execution time reflecting a stable memory consumption of 77.6 MB. The memory consumption of the original SCoPE implementation peaked at 84.9 MB, with a more consistent usage of 80.4 MB for the majority of the process. It is important to note that the majority of the memory consumption is related to the Python process and the loading of the dataset into memory. To further assess the impact of the dataset loading on memory usage, a separate Python session was initiated, in which the same subset of 1000 entries was loaded. This operation alone consumed 71.6 MB of RAM, suggesting that much of the memory usage observed in the experiment is due to the dataset loading process.

\subsection{LLaMA 3.2 Results}
\label{sec:results}

To determine whether overcoming SCoPE's limitations could enhance classification performance, the results were compared with those from previous work~\cite{gonçalves2025evaluatingllama32software}. Table~\ref{tab:results_diversevul} summarizes the findings, showing that the application of SCoPE2 slightly improved the model's performance. This is likely caused by the inclusion of more diverse data, since SCoPE2 is not limited regarding macros usage in code. Since macros are widely used in real-world C/C++ projects to define constants, simplify complex expressions, and improve code reusability, their exclusion in SCoPE could potentially result in an incomplete or less realistic dataset.

\begin{table}[h!]
    \centering
    \caption{Performance using the two SCoPE versions.}
    \label{tab:results_diversevul}
    \begin{tabular}{l@{\hskip 0.5cm}c@{\hskip 0.5cm}c@{\hskip 0.5cm}c@{\hskip 0.5cm}c}
    \hline
    \textbf{Pre-processing} & \textbf{Accuracy} & \textbf{Precision} & \textbf{Recall} & \textbf{F1-Score} \\ \hline
    SCoPE \cite{gonçalves2025evaluatingllama32software} & 66\% & 65\% & 67\% & 66\% \\
    SCoPE2 & 67\% & 67\% & 67\% & 67\% \\ \hline
    \end{tabular}
\end{table}

The most significant improvement was in Precision, with an increase of 2\%. This shows that the application of SCoPE2 lead to an improvement of model's ability to more accurately identify and classify vulnerable code segments, thereby reducing the number of false positives.

\subsection{Threats to Validity}

The potential risks to the validity of the results primarily stem from the data used to train the models. Even though a refined version of DiverseVul was employed, the dataset might still contain some undetected erroneous data points. Furthermore, this study did not investigate whether the chosen LLM had already been pre-trained on data included in the fine-tuning dataset, which could negatively impact its performance.

\section{Conclusion} 
\label{sec:conclusion}

This study introduced SCoPE2, a more efficient and flexible framework for code processing. Compared to the original SCoPE implementation, SCoPE2 delivers substantial performance improvements, reducing execution time by 97.3\%. While the reduction in memory usage was less pronounced, it still marked an improvement. These optimizations are particularly important for processing large datasets. For instance, when processing the entire DiverseVul dataset, SCoPE2 reduced execution time from over 24 hours to just 20 minutes.

By overcoming the limitations of its predecessor, SCoPE2 was successfully applied to process a refined subset of the DiverseVul dataset, previously used in related work to create AI models for SVD. The processed data was then used to fine-tune the LLaMA 3.2 model for vulnerability detection, leading to performance improvements: a 2\% increase in Precision, 1\% in Accuracy, and 1\% in F1-Score compared to results obtained with the original SCoPE framework. SCoPE2 achieved these results with greater efficiency, demonstrating its potential for large-scale source code processing.

The improved efficiency and flexibility of SCoPE2 pave the way for future applications in AI source code analysis tasks. Future work may explore newer AI models for SVD, as well as the usage of SCoPE2 in adversarial example generation and data augmentation tasks.

\begin{credits}
\subsubsection{\ackname}  This work was done and funded in the scope of the BEHAVIOR project  (NORTE2030-FEDER-00576300 no. 14391). This work was also supported by UIDB/00760/2020.

\subsubsection{\discintname}
The authors have no competing interests to declare that are relevant to the content of this article. 
\end{credits}

\printbibliography[heading=bibintoc]

@InProceedings{gonçalves2024scopeevaluatingllmssoftware,
author="Gon{\c{c}}alves, Jos{\'e}
and Dias, Tiago
and Maia, Eva
and Pra{\c{c}}a, Isabel",
editor="Mehmood, Rashid
and Hern{\'a}ndez, Guillermo
and Pra{\c{c}}a, Isabel
and Wikarek, Jaroslaw
and Loukanova, Roussanka
and Monteiro dos Reis, Ars{\'e}nio
and Skarmeta, Antonio
and Lombardi, Eleonora",
title="SCoPE: Evaluating LLMs for Software Vulnerability Detection",
booktitle="Distributed Computing and Artificial Intelligence, Special Sessions I, 21st International Conference",
year="2025",
publisher="Springer Nature Switzerland",
address="Cham",
pages="34--43",
isbn="978-3-031-76459-2",
doi="10.1007/978-3-031-76459-2_4"
}

@article{Razuvayevskaya2023Comparison,
doi = {10.1371/journal.pone.0301738},
    author = {Razuvayevskaya, Olesya AND Wu, Ben AND Leite, João A. AND Heppell, Freddy AND Srba, Ivan AND Scarton, Carolina AND Bontcheva, Kalina AND Song, Xingyi},
    journal = {PLOS ONE},
    publisher = {Public Library of Science},
    title = {Comparison between parameter-efficient techniques and full fine-tuning: A case study on multilingual news article classification},
    year = {2024},
    month = {05},
    volume = {19},
    pages = {1-26},
    number = {5},
}

@ARTICLE{10531717,
  author={Gupta, Aditi and Goyal, Rinkaj},
  journal={IEEE Access}, 
  title={A Generative AI-Driven Method-Level Semantic Clone Detection Based on the Structural and Semantical Comparison of Methods}, 
  year={2024},
  volume={12},
  number={},
  pages={70773-70791},
  doi={10.1109/ACCESS.2024.3401770}}

@article{10.1145/3428230,
author = {Yefet, Noam and Alon, Uri and Yahav, Eran},
title = {Adversarial examples for models of code},
year = {2020},
issue_date = {November 2020},
publisher = {Association for Computing Machinery},
address = {New York, NY, USA},
volume = {4},
number = {OOPSLA},
doi = {10.1145/3428230},
journal = {Proc. ACM Program. Lang.},
month = nov,
articleno = {162},
numpages = {30}
}

@article{10.1145/3630010,
author = {Yang, Guang and Zhou, Yu and Yang, Wenhua and Yue, Tao and Chen, Xiang and Chen, Taolue},
title = {How Important Are Good Method Names in Neural Code Generation? A Model Robustness Perspective},
year = {2024},
issue_date = {March 2024},
publisher = {Association for Computing Machinery},
address = {New York, NY, USA},
volume = {33},
number = {3},
issn = {1049-331X},
doi = {10.1145/3630010},
journal = {ACM Trans. Softw. Eng. Methodol.},
month = mar,
articleno = {60},
numpages = {35}
}

@inproceedings{10.1145/3611643.3616356,
author = {Du, Xiaohu and Wen, Ming and Wei, Zichao and Wang, Shangwen and Jin, Hai},
title = {An Extensive Study on Adversarial Attack against Pre-trained Models of Code},
year = {2023},
isbn = {9798400703270},
publisher = {Association for Computing Machinery},
address = {New York, NY, USA},
doi = {10.1145/3611643.3616356},
booktitle = {Proceedings of the 31st ACM Joint European Software Engineering Conference and Symposium on the Foundations of Software Engineering},
pages = {489–501},
numpages = {13},
location = {San Francisco, CA, USA},
series = {ESEC/FSE 2023}
}

@article{KALOUPTSOGLOU2023107303,
title = {Software vulnerability prediction: A systematic mapping study},
journal = {Information and Software Technology},
volume = {164},
pages = {107303},
year = {2023},
issn = {0950-5849},
doi = {10.1016/j.infsof.2023.107303},
author = {Ilias Kalouptsoglou and Miltiadis Siavvas and Apostolos Ampatzoglou and Dionysios Kehagias and Alexander Chatzigeorgiou},
}

@article{LOMIO2022111283,
title = {Just-in-time software vulnerability detection: Are we there yet?},
journal = {Journal of Systems and Software},
volume = {188},
pages = {111283},
year = {2022},
issn = {0164-1212},
doi = {10.1016/j.jss.2022.111283},
author = {Francesco Lomio and Emanuele Iannone and Andrea {De Lucia} and Fabio Palomba and Valentina Lenarduzzi},
}

@INPROCEEDINGS{10172650,
  author={Croft, Roland and Babar, M. Ali and Kholoosi, M. Mehdi},
  booktitle={2023 IEEE/ACM 45th International Conference on Software Engineering (ICSE)}, 
  title={Data Quality for Software Vulnerability Datasets}, 
  year={2023},
  volume={},
  number={},
  pages={121-133},
  doi={10.1109/ICSE48619.2023.00022}}

@article{LU2024112031,
title = {GRACE: Empowering LLM-based software vulnerability detection with graph structure and in-context learning},
journal = {Journal of Systems and Software},
volume = {212},
pages = {112031},
year = {2024},
issn = {0164-1212},
doi = {10.1016/j.jss.2024.112031},
author = {Guilong Lu and Xiaolin Ju and Xiang Chen and Wenlong Pei and Zhilong Cai},
}

@INPROCEEDINGS{10301302,
  author={Purba, Moumita Das and Ghosh, Arpita and Radford, Benjamin J. and Chu, Bill},
  booktitle={2023 IEEE 34th International Symposium on Software Reliability Engineering Workshops (ISSREW)}, 
  title={Software Vulnerability Detection using Large Language Models}, 
  year={2023},
  volume={},
  number={},
  pages={112-119},
  doi={10.1109/ISSREW60843.2023.00058}}

@INPROCEEDINGS{10774025,
  author={Mahyari, Andrew A.},
  booktitle={MILCOM 2024 - 2024 IEEE Military Communications Conference (MILCOM)}, 
  title={Harnessing the Power of LLMs in Source Code Vulnerability Detection}, 
  year={2024},
  volume={},
  number={},
  pages={251-256},
  doi={10.1109/MILCOM61039.2024.10774025}}

@inproceedings{10114536394763639762,
author = {Zhou, Xin and Zhang, Ting and Lo, David},
title = {Large Language Model for Vulnerability Detection: Emerging Results and Future Directions},
year = {2024},
isbn = {9798400705007},
publisher = {Association for Computing Machinery},
address = {New York, NY, USA},
doi = {10.1145/3639476.3639762},
booktitle = {Proceedings of the 2024 ACM/IEEE 44th International Conference on Software Engineering: New Ideas and Emerging Results},
pages = {47–51},
numpages = {5},
location = {Lisbon, Portugal},
series = {ICSE-NIER'24}
}

@ARTICLE{10879492,
  author={Tamberg, Karl and Bahsi, Hayretdin},
  journal={IEEE Access}, 
  title={Harnessing Large Language Models for Software Vulnerability Detection: A Comprehensive Benchmarking Study}, 
  year={2025},
  volume={13},
  number={},
  pages={29698-29717},
  doi={10.1109/ACCESS.2025.3541146}}

@misc{gonçalves2025evaluatingllama32software,
      title={Evaluating LLaMA 3.2 for Software Vulnerability Detection}, 
      author={José Gonçalves and Miguel Silva and Bernardo Cabral and Tiago Dias and Eva Maia and Isabel Praça and Ricardo Severino and Luís Lino Ferreira},
      year={2025},
      eprint={2503.07770},
      archivePrefix={arXiv},
      primaryClass={cs.LG},
      doi={10.48550/arXiv.2503.07770}, 
}

@article{Yu2023,
  title   = {AdVulCode: Generating Adversarial Vulnerable Code against Deep Learning-Based Vulnerability Detectors},
  author  = {Xueqi Yu and Zhen Li and Xiang Huang and Shasha Zhao},
  year    = 2023,
  month   = 2,
  journal = {Electronics},
  volume  = 12,
  pages   = 936,
  doi     = {10.3390/electronics12040936},
  issn    = {2079-9292},
  issue   = 4
}

@inproceedings{Chen2023,
  series = {RAID 2023},
  title = {DiverseVul: A New Vulnerable Source Code Dataset for Deep Learning Based Vulnerability Detection},
  year = {2023},
  DOI = {10.1145/3607199.3607242},
  booktitle = {Proceedings of the 26th International Symposium on Research in Attacks,  Intrusions and Defenses},
  publisher = {ACM},
  author = {Chen,  Yizheng and Ding,  Zhoujie and Alowain,  Lamya and Chen,  Xinyun and Wagner,  David},
  collection = {RAID 2023}
}

@ARTICLE{10908394,
  author={Shestov, Aleksei and Levichev, Rodion and Mussabayev, Ravil and Maslov, Evgeny and Zadorozhny, Pavel and Cheshkov, Anton and Mussabayev, Rustam and Toleu, Alymzhan and Tolegen, Gulmira and Krassovitskiy, Alexander},
  journal={IEEE Access}, 
  title={Finetuning Large Language Models for Vulnerability Detection}, 
  year={2025},
  volume={13},
  number={},
  pages={38889-38900},
  doi={10.1109/ACCESS.2025.3546700}}

@dataset{goncalves_2025_15051277,
  author       = {Gonçalves, José and
                  Silva, Miguel and
                  Cabral, Bernardo and
                  Dias, Tiago Fontes and
                  Maia, Eva and
                  Praça, Isabel and
                  Severino, Ricardo and
                  Lino Ferreira, Luis},
  title        = {RDiverseVul: Refined DiverseVul},
  month        = feb,
  year         = 2025,
  publisher    = {Zenodo},
  version      = {1.0.0},
  doi          = {10.5281/zenodo.15051277},
}

\end{document}